\documentclass[superscriptaddress,aps,pra,twocolumn,floatfix]{revtex4-2}

\usepackage{graphicx}
\usepackage{dcolumn}
\usepackage{bm}
\usepackage{multirow}%
\usepackage{amsmath,amssymb,amsfonts}%
\usepackage{amsthm}%
\usepackage{mathrsfs}%
\usepackage{bbm}%
\usepackage{stackengine}
 \usepackage[title]{appendix}%
 \usepackage{xcolor}%
 \usepackage{textcomp}%
 \usepackage{booktabs}%
 \usepackage{algorithm}%
 \usepackage{algorithmicx}%
 \usepackage{algpseudocode}%
 \usepackage{listings}%
\usepackage{hyperref}%
\usepackage{xurl}
\usepackage{color}
\definecolor{mygreen}{rgb}{0,0.5,0}
\definecolor{mygrey}{rgb}{0.5,0.5,0.5}
\definecolor{myred}{rgb}{0.75,0,0}
\definecolor{myblue}{rgb}{0,0,0.75}
\definecolor{mymagenta}{cmyk}{0,1,0,0.12}
\definecolor{mycyan}{cmyk}{1,0,0,0.12}
\definecolor{myorange}{rgb}{1.,0.5,0}
\definecolor{myviolet}{rgb}{0.6,0.15,0.6}
\definecolor{mybrown}{cmyk}{0,0.50,1,0.41}
\usepackage[normalem]{ulem}

\newcommand{\bF}{\mathbf{F}}
\newcommand{\bB}{\mathbf{B}}

\newcommand{\bN}{\mathbf{N}}

\newcommand{\Fmax}{{F_\mathrm{max}}}

\newcommand{\subdc}{_\mathrm{dc}}
\newcommand{\subrf}{_\mathrm{rf}}
\newcommand{\bdc}{B\subdc}

\newcommand{\bvdc}{\mathbf{B}\subdc}

\newcommand{\brf}{B\subrf}
\newcommand{\bvrf}{\mathbf{B}\subrf}

\newcommand{\supzero}{^{(0)}}

\newcommand{\dc}{dc}
\newcommand{\rf}{rf}

\newcommand{\omegaL}{\omega_\mathrm{L}}

\renewcommand{\paragraph}{\section}
\renewcommand{\paragraph}[1]{~\\ \noindent \textit{#1} --
}

\usepackage{siunitx}
\DeclareSIUnit\torr{Torr}
\DeclareSIUnit\amagat{amg}
\begin{document}

\title[Article Title]{Quantum noise in a squeezed-light-enhanced multiparameter quantum sensor}

\newcommand{\ICFO}{ICFO - Institut de Ci\`encies Fot\`oniques, The Barcelona Institute of Science and Technology, 08860 Castelldefels (Barcelona), Spain}
\newcommand{\ICREA}{ICREA - Instituci\'{o} Catalana de Recerca i Estudis Avan{\c{c}}ats, 08010 Barcelona, Spain}

\author{Aleksandra Sierant}
\email{aleksandra.sierant@icfo.eu}
\affiliation{\ICFO}

 \author{Diana Méndez-Avalos}
 \affiliation{\ICFO}

 \author{Santiago Tabares Giraldo}
 \affiliation{\ICFO}

\author{Morgan W. Mitchell}
\affiliation{\ICFO}
\affiliation{\ICREA}

\begin{abstract}
We study quantum enhancement of sensitivity using squeezed light in a multi-parameter quantum sensor, the hybrid dc-rf optically pumped magnetometer (hOPM) [Phys. Rev. Applied 21, 034054, (2024)]. Using a single spin ensemble, the hOPM  acquires both the dc field strength (scalar magnetometry), and resonantly detects one quadrature of the ac magnetic field at a chosen frequency (rf magnetometry). In contrast to the Bell-Bloom scalar magnetometer [Phys. Rev. Lett. 127, 193601 (2021)], the back-action evasion in the hOPM is incomplete, leading to a nontrivial interplay of the three quantum noise sources in this system: photon shot noise, spin projection noise, and measurement back-action noise. We observe these interactions using squeezed light as a tool to control the distribution of optical quantum noise between $S_2$ and $S_3$ polarization Stokes components, and the resulting effect on readout quantum noise and measurement back-action. These results demonstrate quantum-enhanced sensitivity in a continuously operating multi-parameter sensor and reveal fundamental trade-offs between sensitivity, back-action, and bandwidth.
\end{abstract}

\maketitle

Quantum sensing of a single parameter (e.g. the scalar magnetic field strength), described by the theory of quantum parameter estimation \cite{Helstrom1969}, has been extensively investigated both theoretically and experimentally \cite{Degen2017, PezzeRMP2018}. Multi-parameter quantum sensing, in contrast, aims to simultaneously estimate multiple parameters \cite{SzczykulskaAPX2016,Demkowicz2020}, with the potential to provide better precision than when estimating each parameter separately. This potential is supported both theoretically \cite{RehacekPRA2017, ProctorPRL2018, CarolloJSMTE2019, LiuJPAMT2020, Gorecki2022} and by proof-of-principle demonstrations \cite{ColangeloN2017, MollerN2017, PolinoO2019}.

Many practical quantum sensors operate continuously, relying on non-destructive monitoring of evolving quantum systems \cite{LIGO2013, Sheng2013}. In optically detected spin systems such as optically-pumped magnetometers (OPMs), three primary quantum noise sources determine the fundamental sensitivity of the sensor: spin projection noise (SPN), photon shot noise (PSN), and measurement back-action (MBA) noise, i.e., the perturbation of the quantum state due to the measurement itself \cite{BraginskyBook1995, Troullinou2021, GanapathyPRX2023}. Approaches to surpass the standard quantum limit (SQL) include optical squeezing \cite{troullinou2021squeezed,bai2021quantum}, spin squeezing \cite{Sewell2012,Colangelo2017}, and techniques to evade or mitigate MBA \cite{Shah2010,Colangelo2017,MartinPRL2017,MollerN2017}.

As we show, the interplay between these quantum noise sources leads to complex quantum limits in continuously operating sensors. In particular, a single, quantum-limited sensor can show a trade-off between sensitivity and bandwidth, and between sensitivity to signals at different frequencies. In this work, we study this interplay using the hOPM \cite{lipka2024multiparameter}, a practical platform for multiparameter quantum estimation. The hOPM employs a single atomic spin ensemble to simultaneously perform scalar dc magnetometry and resonant rf magnetometry -- that is, it measures both the static magnetic field strength and one quadrature of an ac magnetic field at a chosen frequency. Unlike the Bell-Bloom scalar magnetometer (BBOPM) \cite{troullinou2021squeezed}, the hOPM does not evade MBA, making it an ideal testbed for studying the interactions among the three quantum noise sources.

We use polarization-squeezed probe light to control and probe this noise interplay. By adjusting the distribution of quantum fluctuations between the $S_2$ and $S_3$ Stokes components of the probe, we influence both the readout noise and the back-action imprinted on the spin ensemble. Beyond serving as a testbed for noise redistribution, this approach allows us to demonstrate quantum enhancement in a multi-parameter sensor: reducing total quantum noise below the SQL for both dc and rf measurements, and simultaneously improving magnetic sensitivity and measurement bandwidth. The techniques are compatible with emerging technologies for  sensor miniaturization, including integrated OPMs \cite{KitchingAPR2018} and on-chip squeezed-light sources \cite{StefszkyOE2023, ParkSA2024}, suggesting a pathway toward compact, quantum-enhanced MPQS platforms for real-world applications.

\newcommand{\justifying}{}

\begin{figure}[t]
    \centering
    \includegraphics[width=\columnwidth]{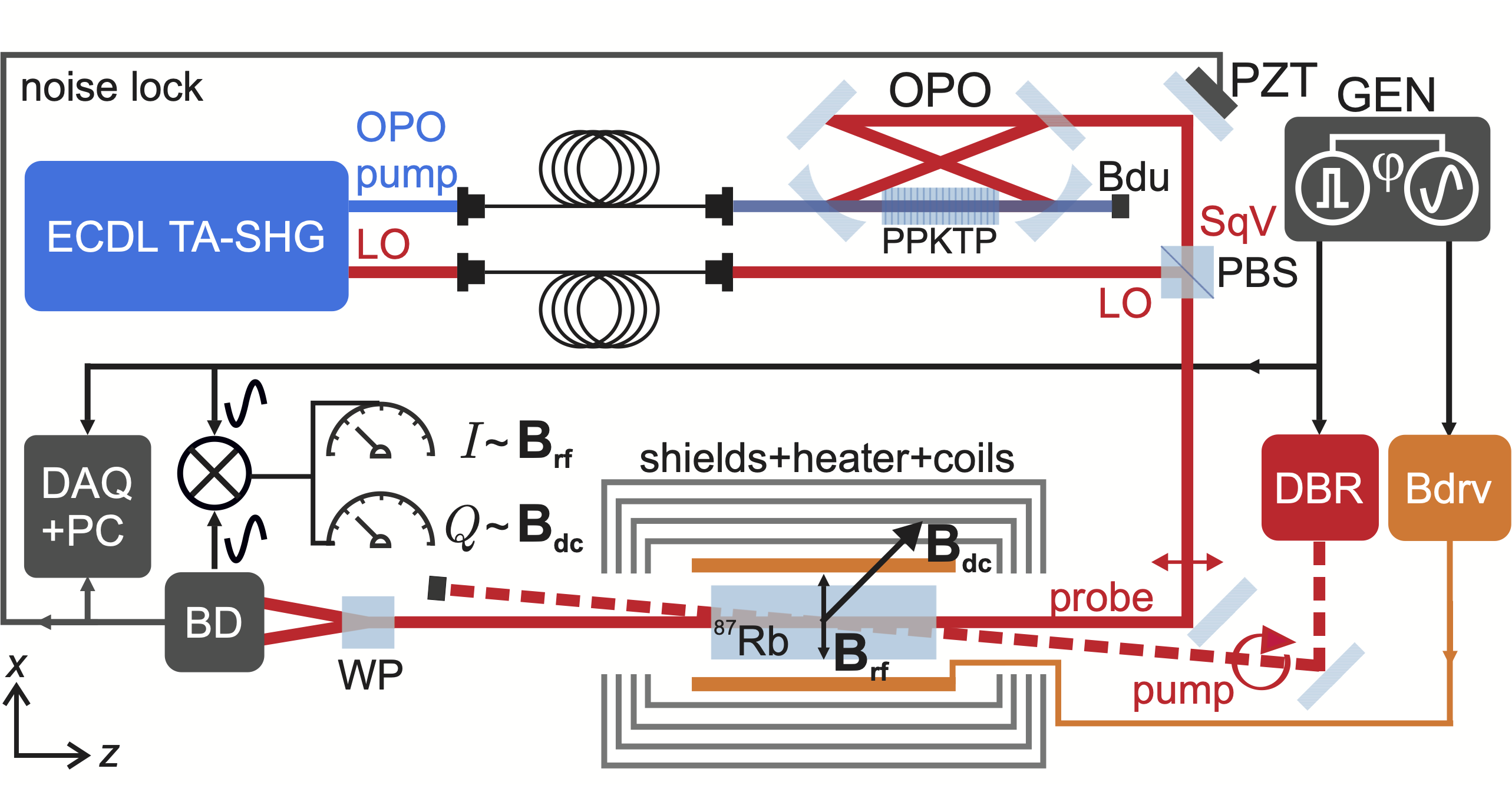}
    \caption{\justifying
    \textbf{Experimental setup of a quantum-enhanced hybrid \dc/\rf{} optically pumped magnetometer (QE-hOPM).}
    WP – Wollaston prism, BD – balanced detector, OPO – optical parametric oscillator, LO – local oscillator, SqV – squeezed vacuum, PM – polarization maintaining fiber, PBS – polarizing beam splitter, DAQ – data acquisition card, Bdrv – low noise current driver, ECDL – extended cavity diode laser, TA-SHG – tapered amplified second harmonic generator, PPKTP – periodically poled potassium titanyl phosphate nonlinear crystal, Bdu – beam dump, GEN – function generator, PZT – piezoelectric element. Description in the text.}
    \label{fig:setup}
\end{figure}

\paragraph{QE-hOPM operation}
The quantum-enhanced hybrid optically pumped magnetometer (QE-hOPM) is shown schematically in Fig.~\ref{fig:setup}. The hOPM simultaneously performs scalar dc magnetometry and resonant rf magnetometry using a single atomic spin ensemble \cite{lipka2024multiparameter}. The static magnetic field $\bvdc$ is inferred from the atomic spin precession (Larmor) frequency $\omega_L$, while one quadrature of an oscillating magnetic field $\bvrf$ is inferred from the amplitude of the spin precession.

The sensor operates in the Bell--Bloom regime \cite{bell1961optically}, with periodic optical pumping at frequency $\omega_{\mathrm{p}} \approx \omega_L$. After interaction with the atomic ensemble, the probe polarization undergoes Faraday rotation proportional to the spin component $F_z$ and is detected with a balanced polarimeter. Demodulation at $\omega_{\mathrm{p}}$ yields the in-phase ($I$) and quadrature ($Q$) components of the signal. The $Q$ quadrature encodes the phase of the spin precession relative to the drive and is used to infer $\bvdc$, while the $I$ quadrature is proportional to the spin precession amplitude and responds to one quadrature of the $\bvrf$ field\cite{lipka2024multiparameter}.

Polarization-squeezed probe light is employed to reduce photon shot noise in the detected Stokes component. The redistribution of quantum fluctuations between conjugate Stokes components modifies both the optical readout noise and the measurement back-action acting on the atomic spin ensemble, enabling controlled exploration of quantum noise trade-offs in a multi-parameter sensor. Experimental parameters and technical details are provided in Appendix \ref{sec:experiment}.

\paragraph{Quantum noise reduction in QE-hOPM}
The quantum noise dynamics of Bell--Bloom magnetometers have been studied in detail in Ref.~\cite{troullinou2021squeezed}, and quantum-noise-limited operation of the hOPM has been demonstrated in Ref.~\cite{lipka2024multiparameter}. Optical probing measures the Stokes parameter
\[
S_2^{\mathrm{(out)}}(t)=S_1^{\mathrm{(in)}}\sin\phi(t)+S_2^{\mathrm{(in)}}\cos\phi(t),
\]
which for small polarization rotation angles $\phi(t)=G F_z(t)$ reduces to \cite{Kong2020,lipka2024multiparameter}
\begin{equation}
S_2^{\mathrm{(out)}}(t)=G F_z(t) S_1^{\mathrm{(in)}} + N_{S_2}(t).
\label{eq:S2Out}
\end{equation}
Here $S_\alpha$ ($\alpha\in\{1,2,3\}$) are the Stokes parameters before and after the atoms, $G$ is a detuning-dependent coupling factor, and $N_{S_2}(t)$ denotes the polarization noise of the detected Stokes component.

The measured spin projection $F_z(t)$ oscillates at the pump frequency $\omega_p$ and relaxes due to depolarization mechanisms, which also generate broadband spin noise as required by the fluctuation--dissipation theorem \cite{aleksandrov1981magnetic,muller2010semiconductor,troullinou2021squeezed}. Expressing the detected signal in terms of slowly varying quadratures yields
\[
S_2^{\mathrm{(out)}}(t)=I\cos(\omega_p t)+Q\sin(\omega_p t).
\]
The demodulation phase is chosen such that $Q=0$ at resonance ($\omega_p=\omega_L$). Near resonance, $Q$ responds linearly to changes in the dc magnetic field, while $I$ responds linearly to changes in the quadrature amplitude of the rf magnetic field \cite{lipka2024multiparameter}.

Single-sided power spectral densities $\mathcal{S}_{I,Q}(\omega)$ are computed using a discrete Fourier transform with a Hann window \cite{troullinou2021squeezed,TroullinouPRL2023}. The measured spectra are well described by
\begin{equation}
\mathcal{S}_{I,Q}(\omega)=\xi^2\mathcal{S}^{\mathrm{PSN}}_{I,Q}
+\mathcal{L}(\omega)\!\left(\mathcal{S}^{\mathrm{SPN}}_{I,Q}
+\bar{\xi}^2\mathcal{S}^{\mathrm{MBA}}_{I,Q}\right),
\label{eq:NoiseSpectra}
\end{equation}
where $\xi^2$ ($\bar{\xi}^2$) is the squeezing (antisqueezing) factor. Photon shot noise (PSN) is frequency independent, while spin projection noise (SPN) and measurement back-action (MBA) exhibit a Lorentzian envelope
$\mathcal{L}(\omega)=\Delta\omega^2/(\omega^2+\Delta\omega^2)$, with $\Delta\omega$ the magnetic-resonance linewidth.

\begin{figure*}[t]
    \centering
            \includegraphics[width=\linewidth]{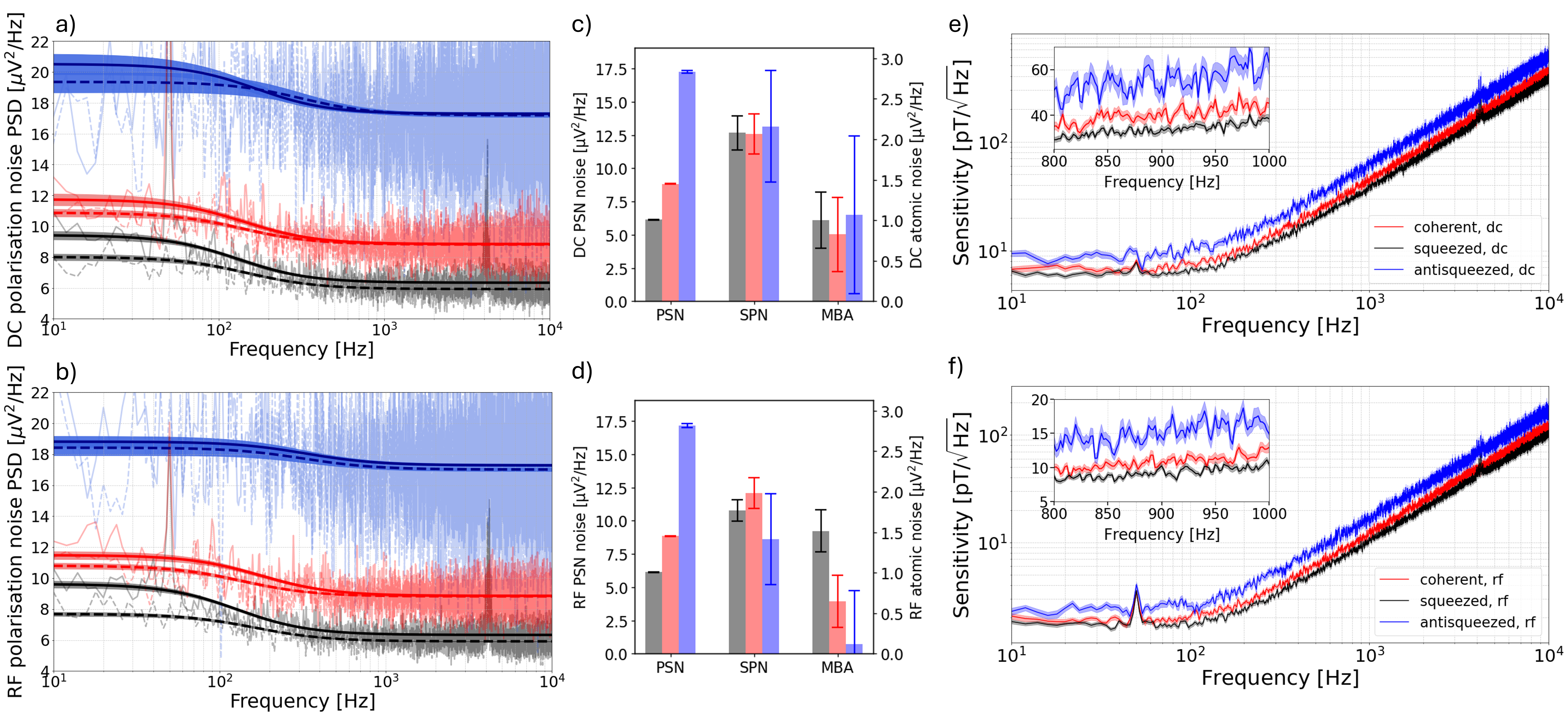}
   \caption{\textbf{Observed quantum noise in the dc (a,c,e) and rf (b,d,f) magnetometer signals for squeezed (black) coherent-state (red) and anti-squeezed (blue) probing.} (a, b) Polarization noise after demodulation. Light traces show data, dark traces show maximum likelihood fits with \autoref{eq:NoiseSpectra}. Probe power \SI{560}{\micro\watt} throughout. Pump power  \SI{110}{\micro\watt} ( solid curves),  \SI{0}{\micro\watt} (spin-noise spectroscopy, dashed curves).  Shaded regions show $\pm 1 \sigma$ uncertainty from bootstrapping. (c, d) best-fit parameters $\mathcal{S}^{\mathrm{PSN}}$, $\mathcal{S}^{\mathrm{SPN}}$, $\mathcal{S}^{\mathrm{MBA}}$. Error bars show the statistical uncertainties of the extracted PSN, SPN, and MBA amplitudes, obtained from the bootstrap-estimated standard deviations of the fitted parameters.  (e,f)  Sensitivity of the hOPM for the dc and rf detection. Shaded regions show $\pm 1 \sigma$ uncertainty from bootstrapping.}
    \label{fig:polNoise}
\end{figure*}

Figure~\ref{fig:polNoise}(a--d) shows the measured polarization noise spectra together with fits to Eq.~\ref{eq:NoiseSpectra}. The data were averaged over 50--150 iterations. Technical noise peaks near \SI{50}{\hertz} and \SI{4}{\kilo\hertz} were excluded from the fitting. Model parameters were extracted following \cite{Lucivero2017a} using maximum-likelihood estimation and bootstrapping to compare the response bandwidth, PSN, and low-frequency atomic noise (SPN+MBA) for coherent, squeezed, and antisqueezed probing, and for polarized and unpolarized atomic ensembles.

The fits show the spectra are dominated by atomic noise (SPN and MBA) at low frequencies and by optical noise (PSN) at high frequencies.  Relative to coherent probing, squeezed light reduces the PSN by approximately \(1.59(1)~\mathrm{dB}\), while antisqueezed probing increases the PSN by about \(2.91(3)~\mathrm{dB}\), with identical values for the dc and rf channels within experimental uncertainty. The SPN remains unchanged within error for all probe states, as expected. In contrast, the MBA contribution increases under squeezed probing, consistent with enhanced quantum back-action accompanying reduced optical shot noise. No systematic MBA reduction is observed for antisqueezed probing. Variations in the polarization-noise bandwidth are observed but remain comparable within uncertainties, indicating that squeezing primarily modifies noise amplitudes rather than the dynamical response.

\paragraph{Sensitivity of QE-hOPM for dc and rf detection}
Magnetic sensitivity of the hOPM is measured as in \cite{lipka2024multiparameter}. The  responsivity, defined as $R_\mathcal{I}(\omega) \equiv d \mathcal{I}(\omega)/ d\brf(\omega)$ ($R_\mathcal{Q}(\omega) \equiv d \mathcal{Q}(\omega)/ d \bdc(\omega)$) is measured in two steps. First, at a range of frequencies $\omega$  we apply a small harmonic perturbation to $\brf$ ($\bdc$) and observe the resulting $S_\mathcal{I}(\omega)$ ($S_\mathcal{Q}(\omega))$. This gives $R^2_\mathcal{I}(\omega)$ ($R^2_\mathcal{Q}(\omega))$ up to a global calibration factor, which we obtain by quasi-statically acquiring $\mathcal{I}$ versus $\brf$  with  $\delta\bdc = 0$ ($\mathcal{Q}$ versus $\bdc$ with $\brf = 0$) and making a linear fit around the operating point to find $R^2_\mathcal{I}(0)$ ($R^2_\mathcal{Q}(0))$.  The equivalent magnetic noise is then calculated as
\begin{eqnarray}
\mathcal{S}_{\brf}(\omega) &=& \mathcal{S}_\mathcal{I}(\omega) / R_\mathcal{I}^{2}(\omega) \\
\mathcal{S}_{\bdc}(\omega) &=& \mathcal{S}_\mathcal{Q}(\omega) / R_\mathcal{Q}^{2}(\omega).
\end{eqnarray}
This method  automatically accounts for frequency dependence of the signal chain and data analysis.
 
 The hOPM is then operated at the nominal operating point with no applied test signal, and the residual noise power spectral density $S_\mathcal{I}(\omega)$ ($S_\mathcal{Q}(\omega)$) is recorded in the same way. 
In the presence of squeezed light, this sensitivity is modified by $\xi^2$ and $\bar \xi^2$, affecting the noise spectrum as given in Eq. \ref{eq:NoiseSpectra}, leaving responsivity unchanged. The measured sensitivity is shown in Fig. \ref{fig:polNoise}. The noise spectra here were averaged over 50-150 iterations and smoothed for clarity using frequency bins whose widths scale proportionally with frequency (logarithmic binning). The sensitivity enhancement provided by squeezed-light probing is seen across both low- and high-frequency bands, see Tab. \ref{tab:sensitivity}. We note that the hOPM is operated in a regime optimized for resolving quantum noise contributions rather than for absolute sensitivity, leading to slightly higher noise floor compared to the sensitivities reported in \cite{lipka2024multiparameter}. The sensitivity of the dc (rf) readout in the PSN dominated region is improved by 1.42(1) dB (1.61(1) dB) for the squeezed-light probing, and reduced by 2.88(2) dB (2.69(2) dB) for the antisqueezed-light probing.
\begin{table}[t]
\centering
\caption{Window-averaged magnetic-field sensitivities for the dc and rf channels using squeezed (sq), coherent (coh), and anti-squeezed (asq) probing. Values in parentheses give the relative noise reduction or penalty in dB with respect to the coherent case. BW is 3~dB bandwidth.}
\label{tab:sensitivity}

\setlength{\tabcolsep}{1pt}
\renewcommand{\arraystretch}{1.15}

\begin{tabular}{c cccc c}
\hline\hline
& \multicolumn{2}{c}{10--80 Hz} & \multicolumn{2}{c}{850--950 Hz} & \\
\cline{2-3}\cline{4-5}
State
& $\mathcal{S}$ ($\si{\pico\tesla\per\sqrt\hertz}$) & dB
& $\mathcal{S}$ ($\si{\pico\tesla\per\sqrt\hertz}$) & dB
& BW (Hz) \\
\hline
$\mathrm{rf}_{\mathrm{sq}}$
& $1.717(2)$ & $-0.88(1)$
& $9.210(8)$  & $-1.42(1)$
& $198$ \\
$\mathrm{rf}_{\mathrm{coh}}$
& $1.901(2)$ & $0$
& $10.842(8)$ & $0$
& $180$ \\
$\mathrm{rf}_{\mathrm{asq}}$
& $2.509(4)$ & $+2.41(2)$
& $15.106(23)$ & $+2.88(2)$
& $144$ \\
\hline
$\mathrm{dc}_{\mathrm{sq}}$
& $6.112(7)$ & $-1.05(1)$
& $33.544(23)$ & $-1.61(1)$
& $164$ \\
$\mathrm{dc}_{\mathrm{coh}}$
& $6.897(11)$ & $0$
& $40.359(36)$ & $0$
& $154$ \\
$\mathrm{dc}_{\mathrm{asq}}$
& $9.167(15)$ & $+2.47(2)$
& $55.011(72)$ & $+2.69(2)$
& $118$ \\
\hline\hline
\end{tabular}
\end{table}

For BBOPM, both the enhancement in high-frequency sensitivity and the broader measurement bandwidth are evident and, in all scenarios, beneficial due to the evasion of MBA -- meaning that low-frequency sensitivity is not sacrificed to achieve a wider bandwidth. In the case of hOPM, high-frequency sensitivity is also improved, along with the slightly increased measurement bandwidth. However, the benefit of the broader bandwidth is not always guaranteed, as it depends on the interplay of various noise contributions. Therefore, the specific frequency range of interest should always be considered to determine whether squeezing provides a net advantage.

\paragraph{MBA in QE-hOPM} 
In the hOPM, the collective spin $\mathbf{F}$ is governed by a Bloch equation for the precession velocity $d\mathbf{F}/dt$ (see Appendix \ref{sec:MBA} for details), to which MBA contributes a term $GS_3\hat z \times \textbf{F}$. Here $GS_3\hat z$ can be interpreted as a fictitious field produced by ac Stark shifts \cite{HapperMathur1967PR} induced by the probe light, i.e., optical Zeeman shifts (OZS). 
\begin{figure}[t]
    \centering
    \includegraphics[width=\columnwidth]{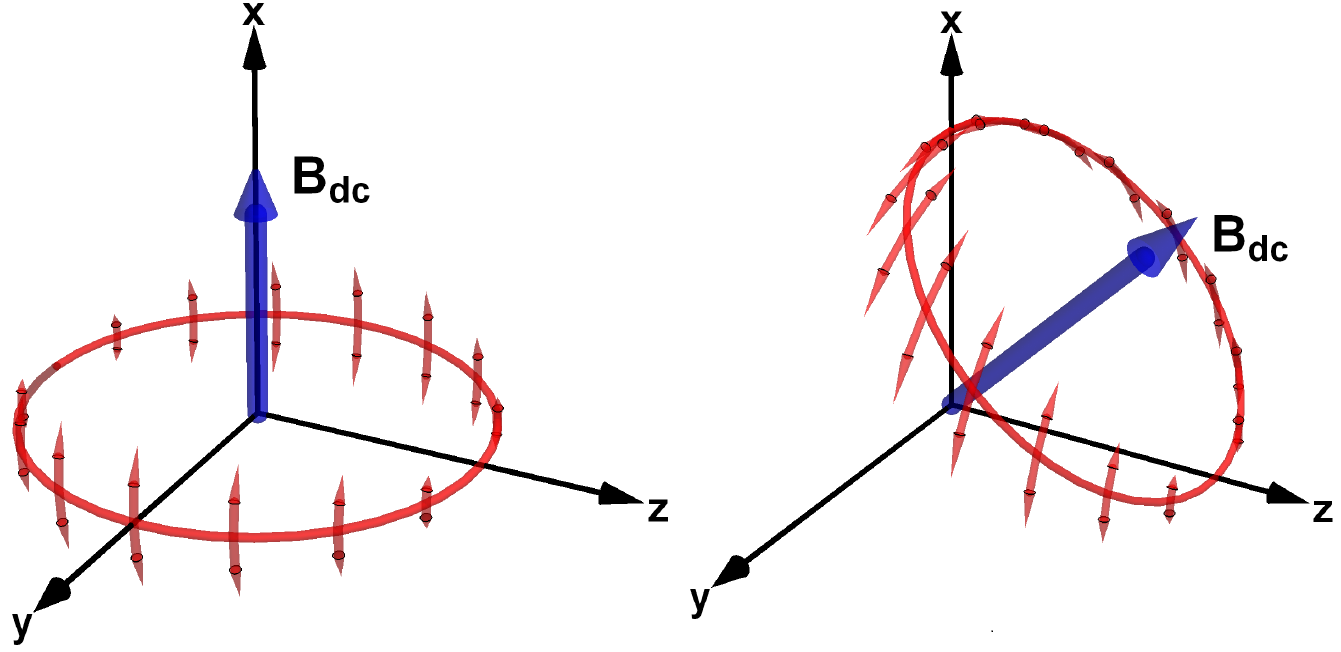}
    \caption{\justifying
    \textbf{Visualization of the effect of $S_3$ noise on spin precession.}
    Left: BBOPM, as in \cite{Troullinou2021,TroullinouPRL2023}. Right: hOPM, as in \cite{lipka2024multiparameter}. The probing direction is along $z$. The thick blue arrow shows the magnetic-field direction. Red double-headed arrows represent perturbations to spin precession (direction and relative magnitude) due to the optical Zeeman shift (OZS), which produces a random rotation about the $z$ axis proportional to the ellipticity, i.e., the Stokes component $S_3$.
    }
    \label{fig:MBAVisualization}
\end{figure}
The OZS noise term $GS_3\hat z$ is weak compared to the bias field that drives the spin precession, and is treated as a perturbation.  Components of this noise near the Larmor frequency accumulate to perturb the spin by an amount comparable to the intrinsic spin variance. Visualization of this effect on spin precession is shown in Fig. \ref{fig:MBAVisualization}, for the BBOPM and hOPM. In the BBOPM, the OZS perturbs in first order only the $F_x$ component, and thus neither the amplitude nor the phase of the spin precession, which is in the y-z plane. This makes the BBOPM MBA-evading.

In the hOPM, the same $GS_3\hat z$ noise term due to OZS affects all of the $F_x$, $F_y$ and $F_z$ components in first order, and thus can change both the amplitude and phase of the spin precession. As a result, hOPM is not measurement backaction evading. This also explains why when using squeezed light -- which reduces noise in $S_2$ and increases noise in $S_3$, as these two components are constrained by an uncertainty relation -- we observe increased low-frequency MBA noise in the measured signal with respect to the MBA contribution observed for coherent light (Fig. \ref{fig:polNoise}). We note that despite this increase, the overall low-frequency noise (MBA + PSN + SPN) remains no worse for the squeezed probing than for the coherent probing.

\paragraph{Conclusions}
We have studied the interplay of photon shot noise, spin projection noise, and measurement back-action in a high-performance multi-parameter quantum sensor. With the help of squeezed probe light, we separate the various noise contributions and observe experimentally that squeezing the $S_2$ Stokes component improves the sensitivity to both slow and rapid changes in measured dc and rf fields. We also find evidence that $S_3$ anti-squeezing, which necessarily accompanies $S_2$ squeezing, increases measurement back-action noise, such that excessive squeezing would worsen the sensitivity to slow changes. These observations imply not a single quantum sensitivity limit, but rather a trade-off between enhancement of different parts of the sensitivity spectrum. Similarly, we observe bandwidth enhancement due to $S_2$ squeezing. A simple model shows that these trade-offs are the result of the multi-parameter operation of the sensor, and are not present in its single-parameter version. These observations are expected to apply to other continuously-operating sensors, including both atomic and opto-mechanical instruments. The present experiment is, to our knowledge, the first in which these effects can be simultaneously observed.

\paragraph{Acknowledgments}
The authors thank Charikleia Troullinou for discussion and useful comments on the manuscript, and Piotr Sierant for his help with data processing. This work was supported by the European Union (ERC, Field-SEER, 101097313),  (QUANTIFY 101135931), European Defense Fund (EDF) under grant agreement EDF-2021-DIS-RDIS-ADEQUADE (101074977), Spanish Ministry of Science MCIN project SAPONARIA (PID2021-123813NB-I00), ``NextGenerationEU/PRTR.'' (Grant FJC2021-047840-I) and ``Severo Ochoa'' Center of Excellence CEX2019-000910-S and CEX2024-001490-S;  Generalitat de Catalunya through the CERCA program,  DURSI grant No. 2021 SGR 01453 and QSENSE (GOV/51/2022).  Fundaci\'{o} Privada Cellex; Fundaci\'{o} Mir-Puig. Funded by the European Union. DMA acknowledges funding from the European Union’s Horizon Europe research and innovation programme under the MSCA Grant Agreement No. 101081441. Views and opinions expressed are however those of the author(s) only
and do not necessarily reflect those of the European Union. Neither
the European Union nor the granting authority can be held responsible for them. \\

\subsection*{End Matter}

\subsubsection{Experimental setup and operating parameters}
\label{sec:experiment}

The QE-hOPM is shown schematically in Fig.~\ref{fig:setup}. The construction of the hOPM and the generation of squeezed probe light are described in detail in Refs.~\cite{lipka2024multiparameter} and \cite{Predojevic2008,troullinou2021squeezed}, respectively.

The dc magnetic field $\bvdc$ is applied in the $x$--$z$ plane at $45^\circ$ with respect to the pumping and probing direction. The rf magnetic field $\bvrf$ is aligned along the $x$ axis and oscillates at frequency $\omega_{\mathrm{rf}}$, which is set equal to and phase synchronized with the pump repetition frequency $\omega_{\mathrm{p}}$. Optical pumping is implemented using the Bell--Bloom technique \cite{bell1961optically}, in which the collective atomic spin $\mathbf{F}$ is driven periodically at frequency $\omega_{\mathrm{p}} \approx \omega_L$.

Isotopically enriched $^{87}$Rb vapor is contained in a magnetically shielded, \SI{3}{\centi\meter}-long glass cell filled with \SI{100}{\torr} of N$_2$ buffer gas (pressure specified at room temperature). The atoms are heated to \SI{105}{\celsius} using intermittent Joule heating, yielding an atom number density of \SI{8.2e12}{atoms\per\centi\meter\cubed}, which is optimal for scalar magnetometry under these conditions \cite{TroullinouPRL2023}. Magnetic fields are generated by induction coils along the $x$ (TwinLeaf CSUA300) and $z$ (TwinLeaf CSBA-10) axes driven by low-noise current sources, producing a dc bias field of $\bvdc \approx \SI{6}{\micro\tesla}$, corresponding to $\omega_L \approx 2\pi \times \SI{42}{\kilo\hertz}$. The rf field is generated by current modulation of the $x$-axis coil at $\omega_{\mathrm{rf}} = \omega_L$.

Optical pumping is provided by a circularly polarized beam with instantaneous power \SI{110}{\micro\watt} from a distributed Bragg reflector (DBR) laser propagating along the $z$ axis. The pump frequency is modulated at $\omega_{\mathrm{p}}$ around the $^{87}$Rb D$_1$ transition, sweeping through resonance twice per spin-precession cycle. Pump modulation is generated by a waveform generator (Siglent SDG1025) with phase and frequency synchronized to the rf drive.

The Faraday rotation of a \SI{560}{\micro\watt} linearly polarized probe beam is detected using a shot-noise-limited polarimeter consisting of a half-wave plate, a Wollaston prism, and a balanced detector (Thorlabs PDB450A). The probe and local oscillator are derived from a frequency-doubled diode laser system with a tapered amplifier (Toptica ECDL TA-SHG 110) operating at a fundamental wavelength of \SI{795}{\nano\meter}. The laser is frequency stabilized \SI{50}{\giga\hertz} blue-detuned from the D$_1$ line using a fiber interferometer \cite{Kong2015}. A fraction of the infrared output serves as the coherent probe or local oscillator, while the remaining light is frequency doubled to \SI{397}{\nano\meter}, spatially mode cleaned, and used to pump a subthreshold optical parametric oscillator, generating vertically polarized squeezed vacuum at \SI{795}{\nano\meter} \cite{Predojevic2008,troullinou2021squeezed}.

The squeezed vacuum is combined with the horizontally polarized, mode-matched local oscillator on a polarizing beam splitter to produce a polarization-squeezed probe beam. The relative phase between the squeezed vacuum and the local oscillator is actively stabilized using a piezoelectric actuator and feedback loop \cite{troullinou2021squeezed}. The measured polarization squeezing is 2.0~dB before the atomic cell and 1.8~dB at the balanced detector after the atoms. Detector signals and the pump current monitor are digitized using a data acquisition card (National Instruments PCI-4462) for digital demodulation. We note that the present experiment is limited by optical losses, residual phase noise, and the available pump power for squeezed-light generation.

\subsubsection{Measurement back-action}
\label{sec:MBA}

\newcommand{\bea}{\begin{eqnarray}}
\newcommand{\eea}{\end{eqnarray}}
\newcommand{\nn}{\nonumber \\ }
\newcommand{\nnp}{\nonumber \\ & & +}
\newcommand{\nnm}{\nonumber \\ & & -}
\newcommand{\nne}{\nonumber \\ & = & }
\newcommand{\nnequiv}{\nonumber \\ & \equiv & }
\newcommand{\subcycle}{_\mathrm{cycle}}
\newcommand{\subtrans}{_\perp}
\newcommand{\subtangent}{_{||}}
\newcommand{\Bhat}{\hat{\mathbf{B}}}

Here we follow the model given in \cite[supplement]{troullinou2021squeezed}, in which the spin evolves as 
\begin{align}
\label{eq:OriginalSpinDynamics}
\frac{d}{dt} \bF(t) 
  &= [-\gamma \bB\subdc(t) + G S_3(t) \hat{z} ] \times \bF(t) 
  - \Gamma  \bF(t) \nonumber \\
  &\quad + P(t)[\hat{z} \Fmax -\bF(t)] + \bN_F (t),
\end{align}
where $\bB\subdc$ is the dc magnetic field oriented along $\Bhat\subdc \equiv \mathbf{B}\subdc/|\mathbf{B}\subdc|$, $G S_3 \hat{z}$ is the OZS-induced effective field responsible for MBA, $\Gamma$ is the relaxation rate, $\Fmax$ is the maximum possible polarization, $P$ is the optical pumping rate and $\bN_F$ is a stochastic term, which accounts for SPN.

The evolving spin $\mathbf{F}(t)=\mathbf{F}\supzero(t)+\alpha \mathbf{F}^{(1)}(t)$ includes a deterministic contribution $\mathbf{F}^{(0)}(t)$, which describes collective spin precession about $\mathbf{B}\subdc$ when driven by optical pumping and subject to isotropic relaxation. We define the transverse component of $\mathbf{F}(t)$, i.e., the part in a plane orthogonal to $\mathbf{B}\subdc$, as
\begin{eqnarray}
\mathbf{F}\subtrans (t) \equiv \mathbf{F} (t) - (\mathbf{F} (t) \cdot \Bhat\subdc) \Bhat\subdc.
\end{eqnarray}
Analogously, the transverse part of the zeroth-order contribution is
\begin{eqnarray}
\mathbf{F}\subtrans\supzero (t) \equiv 
\mathbf{F}\subtrans\supzero (t) - (\mathbf{F}\supzero (t) \cdot  \Bhat\subdc) \Bhat\subdc.
\end{eqnarray}
SPN, MBA, and fluctuations of $\bdc$  contribute stochastic perturbations to $ \mathbf{F}^{(1)}(t)$. In particular,  MBA contributes an additive noise term proportional to $S_3(t) \hat{z} \times \mathbf{F}^{(0)}(t)$. 

For the BBOPM, $\Bhat\subdc = \hat{x}$ and $\mathbf{F}\supzero(t)=
\mathbf{F}
\subtrans\supzero(t)$ is in the $y$-$z$ plane. Consequently, $S_3(t) \hat{z} \times \mathbf{F}^{(0)}(t)$ vanishes and there is zero MBA contribution to $\mathbf{F}^{(1)}\subtrans(t)$. Rather, in first order the MBA is entirely in the $F_x$ component, which however does not contribute to the signal $S_2$. The BBOPM is thus back-action evading. 

In the case of the hOPM, which for generality we consider with the B-field tipped an angle $\psi$ from the $\hat{x}$ axis,
the spin precesses about the $\Bhat\subdc = \hat{x} \cos \psi + \hat{z} \sin \psi$ direction. On resonance, $\omega_\mathrm{p} = \omegaL$, with one brief pump pulse per cycle, and with a high quality-factor oscillation, i.e., with $\omegaL \gg \Gamma$, the unperturbed spin trajectory can be approximated as
\begin{eqnarray}
\mathbf{F}\supzero(t) &\propto& (-\hat{x} \sin\psi + \hat{z} \cos\psi) \cos\omegaL t 
+ \hat{y} \sin\omegaL t 
\nonumber \\ & & 
+ (\hat{x} \cos \psi + \hat{z} \sin \psi) \tan \psi 
\end{eqnarray}
and thus
\begin{eqnarray}
\hat{z} \times \mathbf{F}\supzero(t) 
&\propto& \hat{y} (1 - \cos\omegaL t) \sin \psi - \hat{x} \sin \omegaL t  \hspace{9mm}
\\
\mathbf{F}\subtrans\supzero(t) 
&\propto& 
(-\hat{x} \sin\psi + \hat{z} \cos\psi) \cos\omegaL t  
\nonumber \\ & & + \hat{y} \sin\omegaL t 
\\
\mathbf{F}^{(0)}(t) \times \Bhat\subdc & \propto &   (\hat{x} \sin\psi - \hat{z} \cos\psi) \sin\omegaL t
\nonumber \\ & & + \hat{y} \cos\omegaL t 
\end{eqnarray}

Due to the MBA contribution, the square magnitude  $|\mathbf{F}\subtrans(t)|^2$ and the angle $\theta_{\mathbf{F}\subtrans}(t)$ of the precessing {transverse} spin are perturbed at rates
\begin{eqnarray}
\frac{d}{dt} |\mathbf{F}\subtrans(t)|^2 &\propto &  S_3(t) [\hat{z} \times \mathbf{F}^{(0)}(t)] \cdot \mathbf{F}\subtrans^{(0)}(t)
\nonumber \\ & \propto & 
S_3(t) \sin \omegaL t   \sin \psi
\hspace{9mm}
\label{eq:amp}
\end{eqnarray}
and
\begin{eqnarray}
\frac{d}{dt} \theta_{\mathbf{F}\subtrans}(t) &\propto &  S_3(t) [\hat{z} \times \mathbf{F}^{(0)}(t)] \cdot [\mathbf{F}^{(0)}(t) \times \Bhat\subdc] 
\nonumber \\ & \propto & 
S_3(t) (1-\cos \omegaL t) \sin\psi,
\label{eq:phase}
\end{eqnarray}
respectively.

From \autoref{eq:amp} and \autoref{eq:phase} we see that the amplitude and angle are both perturbed by $S_3(t)$, except at particular phases $\omegaL t$ of the precession cycle. Since $S_3(t)$ is delta-correlated, MBA noise imparted by $S_3(t)$ at one instant will not be canceled by noise imparted by $S_3(t)$ at other instants. Thus the hOPM accumulates measurement back-action in both components of $\mathbf{F}\subtrans$. This MBA noise then manifests in both quadratures of the measured Stokes parameter $S_2(t)$, and thus in the low-frequency parts of both the dc and rf field measurements.  The noise powers in both cases are proportional to $\sin^2 \psi$, offering a convenient ``knob'' by which to adjust the MBA effects.  Only if $\psi = 0$ does the measurement become fully MBA-evading - at which point the OPM is only measuring the dc field; the response to rf fields vanishes. It is precisely the multi-parameter estimation capability that makes the sensor vulnerable to MBA.

%

\end{document}